\documentclass[aps,prd,preprint,showpacs,eqsecnum,superscriptaddress,nofootinbib]{revtex4}

\usepackage{amssymb,graphicx}

\def\fig#1{fig.~{\ref{#1}}}

\begin{document}

\noindent \hfill Brown-HET-1553

\vskip 1 cm

\title{Three-Loop Leading Singularities and BDS Ansatz\\
for
Five Particles}

\author{Marcus Spradlin}

\affiliation{Brown University, Providence, Rhode Island 02912, USA}

\author{Anastasia Volovich}

\affiliation{Brown University, Providence, Rhode Island 02912, USA}

\author{Congkao Wen}

\affiliation{Brown University, Providence, Rhode Island 02912, USA}

\begin{abstract}
We use the leading singularity technique to determine the planar
three-loop five-particle amplitude in ${\cal N}=4$ super Yang-Mills
in terms of a simple basis of integrals. We
analytically compute the integral coefficients
for both the parity-even and the parity-odd parts of the amplitude.
The parity-even part involves only dual conformally invariant integrals.
Using the method of obstructions we numerically evaluate two previously
unfixed coefficients
which appear in the three-loop BDS ansatz.

\end{abstract}

\pacs{11.15.Bt, 12.60.Jv, 11.25.Tq}

\maketitle

\section{Introduction}

Scattering amplitudes in gauge theories are remarkable
objects with many properties hidden in the complexity of their
Feynman diagram expansion.  It is natural to expect the most
symmetric gauge theory, ${\mathcal{N}} = 4$ super-Yang-Mills (SYM),
to be especially rich.  Two very recent
papers~\cite{Berkovits:2008ic,Beisert:2008iq}
have used AdS/CFT to relate and shed light on two
particularly remarkable properties of scattering amplitudes
in SYM.  These properties are dual conformal invariance,
first observed for MHV amplitudes in~\cite{Drummond:2006rz} and
conjectured to extend to all amplitudes in~\cite{Drummond:2008vq},
and an equality between amplitudes and certain lightlike
Wilson loops, first established at strong coupling
in~\cite{AM} and at weak coupling in~\cite{DrummondVanishing,BrandhuberWilson}.
These developments were pushed forward in a number of
papers including~\cite{DHKSTwoloopBoxWilson,ConformalWard,HexagonWilson,Bern:2008ap,SV:2008,Brandhuber:2008pf,Drummond:2008bq}
and have been reviewed in~\cite{Alday:2008yw}, where a
comprehensive
set of references may be found.

Much of the recent interest in multi-loop scattering amplitudes has been
stimulated by
the ABDK/BDS ansatz~\cite{ABDK,BDS} which suggested that
multi-loop MHV amplitudes satisfy a powerful iteration relation implying
a simple exponential form for the full all-loop amplitude.
Although the ABDK/BDS ansatz was successfully tested for four particles at
two~\cite{ABDK} and three~\cite{BDS} loops, as well as for five
particles at two loops~\cite{TwoLoopFiveA,TwoLoopFiveB},
some doubts raised in~\cite{AMTrouble,BNST,Lipatov} necessitated
an explicit calculation of the two-loop six-particle
amplitude~\cite{Bern:2008ap} which conclusively demonstrated the
incompleteness of the BDS ansatz.

Indeed six particles is the earliest that the hypothesized dual conformal
symmetry of amplitudes could have allowed BDS to break down; for $n=4,5$
the symmetry fixes the form of the amplitude up to a few numerical
constants~\cite{AMTrouble,ConformalWard}.   Beginning at $n=6$ dual conformal
invariance becomes substantially weaker, determining the form of the
amplitude only up to an arbitrary function of dual conformally invariant
cross-ratios.  Nevertheless it was found in parallel
work~\cite{Bern:2008ap,SV:2008} that the amplitude/Wilson loop equality
survives at two loops for $n=6$ despite not being required
by dual conformal invariance.

In this paper we study the three-loop BDS ansatz
\begin{equation}
M_n^{(3)}(\epsilon) + \frac{1}{3} (M_n^{(1)}(\epsilon))^3
- M_n^{(1)}(\epsilon) M_n^{(2)}(\epsilon)
- f^{(3)}(\epsilon)
M_n^{(1)}(3 \epsilon) = C^{(3)} + {\cal O}(\epsilon)
\label{eq:bdsthree}
\end{equation}
where $C^{(3)}$ is a previously undetermined numerical constant.
In sections II and III
we use the leading singularity method~\cite{Cachazo:2008vp}
to determine the
(four-dimensional cut-constructible part of the)
3-loop 5-particle amplitude $M_5^{(3)}$ in
terms of a simple basis of integrals.  In section IV
we then numerically evaluate
enough pieces of these amplitudes (the pieces called
`obstructions' in~\cite{Cachazo:2006mq,Cachazo:2006az})
to determine $C^{(3)} = 17.8241$.
Although current developments strongly suggest
that the quantity appearing on the right-hand side of~(\ref{eq:bdsthree})
will in general be non-constant (but still dual conformally invariant)
for $n>5$, there is some utility in knowing the precise
number $C^{(3)} = 17.8241$
since for any $n$, whatever appears on the right-hand
side of~(\ref{eq:bdsthree})
must approach this same number in any collinear limit.

\section{Outline of the Calculation}

Our goal is to find a compact expression for the
planar 3-loop 5-particle amplitude in ${\cal{N}}=4$ SYM as a linear
combination of some basic integrals.
Several powerful and related techniques
for carrying out calculations such as these include
unitarity based
methods~\cite{Bern:1994zx,Bern:1994cg,Bern:1995db,Bern:1996je,BDDKSelfDual,Bern:1997sc,Bern:2004cz}
and more recently, building on~\cite{BCFLoop,Buchbinder:2005wp},
maximal cuts~\cite{Bern:2007ct}
and the leading singularity method~\cite{Cachazo:2008vp}.
For the present calculation we find it convenient to use
the leading singularity
method (see
also~\cite{Cachazo:2008dx,Cachazo:2008hp})
since it allows for all integral coefficients
to be determined analytically by solving
simple linear equations.
In this section we provide a detailed outline of the steps involved
in setting up the calculation.

\subsection{Review of the Leading Singularity Method}

Suppose we are interested in calculating some $L$-loop scattering
amplitude ${ A}$. On the one hand, the amplitude may of course be
represented as a sum over Feynman diagrams $F_j$,
\begin{equation}
\label{eq:one} { A}(k) = \sum_j \int \prod_{a=1}^L d^d \ell_a\
F_j(k,\ell),
\end{equation}
where $k$ are external momenta and $\ell_a$ are the loop momenta.
However it is frequently the case, especially in theories as rich as
${\cal N} = 4$ SYM, that directly calculating the sum over Feynman
diagrams would be impractical. Rather the calculation proceeds by
expressing ${ A}$ as a linear combination of relatively simple
integrals in some appropriate basis $\{I_i\}$,
\begin{equation}
\label{eq:two} { A}(k) = \sum_i c_i(k) \int \prod_{a=1}^L d^d
\ell_a\ I_i(k,\ell),
\end{equation}
and then determining the coefficients $c_i$ by other means.

With the leading singularity method we
equate~(\ref{eq:one}) and~(\ref{eq:two})
and perform the integral
\begin{equation}
\label{eq:method} \sum_i c_i(k) \int_\Gamma d^4 \ell\, I_i(k, \ell)
= \int_\Gamma d^4 \ell\,
 \sum_j F_j(k, \ell)
\end{equation}
over contours $\Gamma \in {\mathbb{C}}^{4 L}$ other than the
real $\ell$-axis.
At $L$ loops each contour is a $T^{4L}$ inside
$\mathbb{C}^{4 L}$.
For each contour $\Gamma$
we obtain one linear equation on the coefficients $c_i$.
Of course if $\Gamma$ is a random contour then we would generally
get the useless equation $0=0$, so we should choose contours such that
the integral on the right-hand side of~(\ref{eq:method}) evaluates the
residue on the isolated singularities of Feynman diagrams, which are
associated with the locus where internal propagators become on-shell.
Since the number of isolated singularities in a
generic $L$-loop diagram can be as high as $2^L$ (simple diagrams
can have fewer isolated singularities), the leading singularity
method gives rise to an exponentially large (in $L$) number of
linear equations for the coefficients $c_i$.
We note that the
homogeneous part of these linear equations (the left-hand side
of~(\ref{eq:method})) depends only on the set of integrals $\{I_i\}$
and the choice of contours, while the details of which particular
helicity configuration is under consideration enters only into the
inhomogeneous terms on the right-hand side.

\subsection{Integration Strategy: Collapse and Expand}

Here we briefly review
from~\cite{Buchbinder:2005wp,Bern:2007ct,Cachazo:2008dx,Cachazo:2008vp,Cachazo:2008hp}
the integration rules which make it
simple to evaluate the contour integrals appearing in~(\ref{eq:method})
in the cases relevant to the present calculation.
Let us focus on a box
with loop momentum $p$ and external momenta $k_i$.
The box may be sitting inside a higher-loop diagram, in which case the
$k_i$ may involve other loop momenta.
The sum over Feynman diagrams contains poles at the locus
\begin{equation}
S = \{ p \in {\mathbb{C}}^4 : p^2 = 0,~(p-k_1)^2=0,~(p-k_{12})^2=0~,
(p+k_4)^2=0\},
\end{equation}
which, for generic $k_i$, consists of two distinct points.
To each of these points there is an associated contour $\Gamma_p$ such
that integrating $p$ over $\Gamma_p$ calculates the residue at the
associated point.

The residue of a one-loop amplitude at one of these
poles is computed by removing the four
internal propagators and evaluating the product of
on-shell tree amplitudes
at the four corners (summed over all helicities of internal
states).    In the simplest application, when all four $k_i$ satisfy
$k_i^2 = 0$, this product evaluates
on either contour $\Gamma_p$
to a four-particle tree amplitude,
leading to the `collapse rule'
graphically depicted as
\begin{equation}
\int_{\Gamma_p} d^4 p
{\hbox{\lower 46.5pt\hbox{
\includegraphics{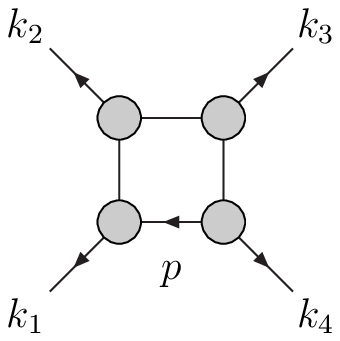}
}}}
=
{\hbox{\lower 36.5pt\hbox{
\includegraphics{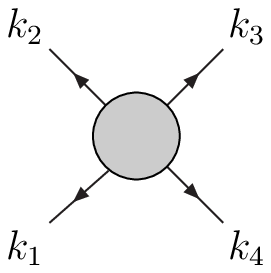}
}}}.
\label{eq:collapse}
\end{equation}
The figure on the left indicates the sum over that subset of
all one-loop Feynman diagrams in which all four of the indicated
propagators are present.  Of course it may as well be
the sum over {\it all} one-loop Feynman diagrams since those
that do not contain all four of the indicated propagators contribute
zero to the residue.

When one of the $k_i^2$ is non-zero, the result~(\ref{eq:collapse})
holds on only one
of the two $\Gamma_p$ contours, while the integral over the other contour
gives zero.
Given a helicity assignment for the external particles it
is a simple matter to determine which of the two solutions leads to
the non-zero result.

It is frequently the case that after collapsing a box in
some loop momentum $p$ there are less than four exposed
propagators in some other loop momentum, which would apparently
indicate a codimension 1 singularity rather than an isolated singularity.
In this case one can use the `expand rule'
\begin{eqnarray}
{\hbox{\lower 36.5pt\hbox{
\includegraphics{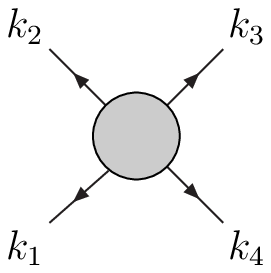}
}}}
&=&
{\hbox{\lower 31.75pt\hbox{
\includegraphics{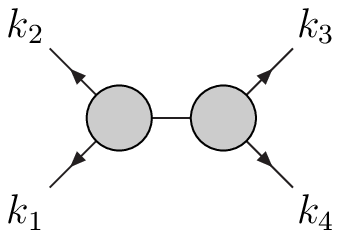}
}}}
+ {\mbox{terms non-singular at~}} (k_1 + k_2)^2 = 0
\cr
&=&
{\hbox{\lower 46.5pt\hbox{
\includegraphics{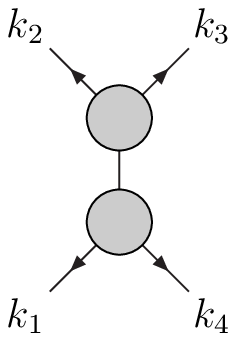}
}}}
+ {\mbox{terms non-singular at~}} (k_2 + k_3)^2 = 0
\nonumber
\end{eqnarray}
to expose additional propagators inside a tree amplitude.
The choice of how to expand is correlated with the choice of integration
contour for the next loop momentum.  In the example shown here, the
terms isolated on the first line are those which survive a contour
integration around the singularity at $(k_1+k_2)^2 = 0$ while the
second expansion displays those terms isolated by a contour integration
around the singularity at $(k_2 + k_3)^2 = 0$.

These two simple rules are sufficient for evaluating all contour
integrals appearing on the right-hand side of~(\ref{eq:method}) in
this paper.
Finally, scalar
integrals appearing on the left-hand side
are integrated via the simple rule
\begin{equation}
\label{eq:jacobian}
\int_{\Gamma_p} d^4 p\,
\frac{1}{p^2 (p - k_1)^2 (p - k_{12})^2 (p + k_4)^2}
= \frac{1}{(k_1 + k_2)^2 (k_2 + k_3)^2},
\end{equation}
which is valid as long as at least three of the $k_i$ satisfy
$k_i^2 = 0$ (we will not encounter any other cases in the present
calculation).
We have chosen a simple normalization factor of $1$ on the right-hand
side of~(\ref{eq:jacobian}); this will be adjusted below in~(\ref{eq:assembly})
to match standard conventions for normalizing amplitudes.

\subsection{Choosing a Sufficient Set of Contours}

\begin{figure}
\includegraphics{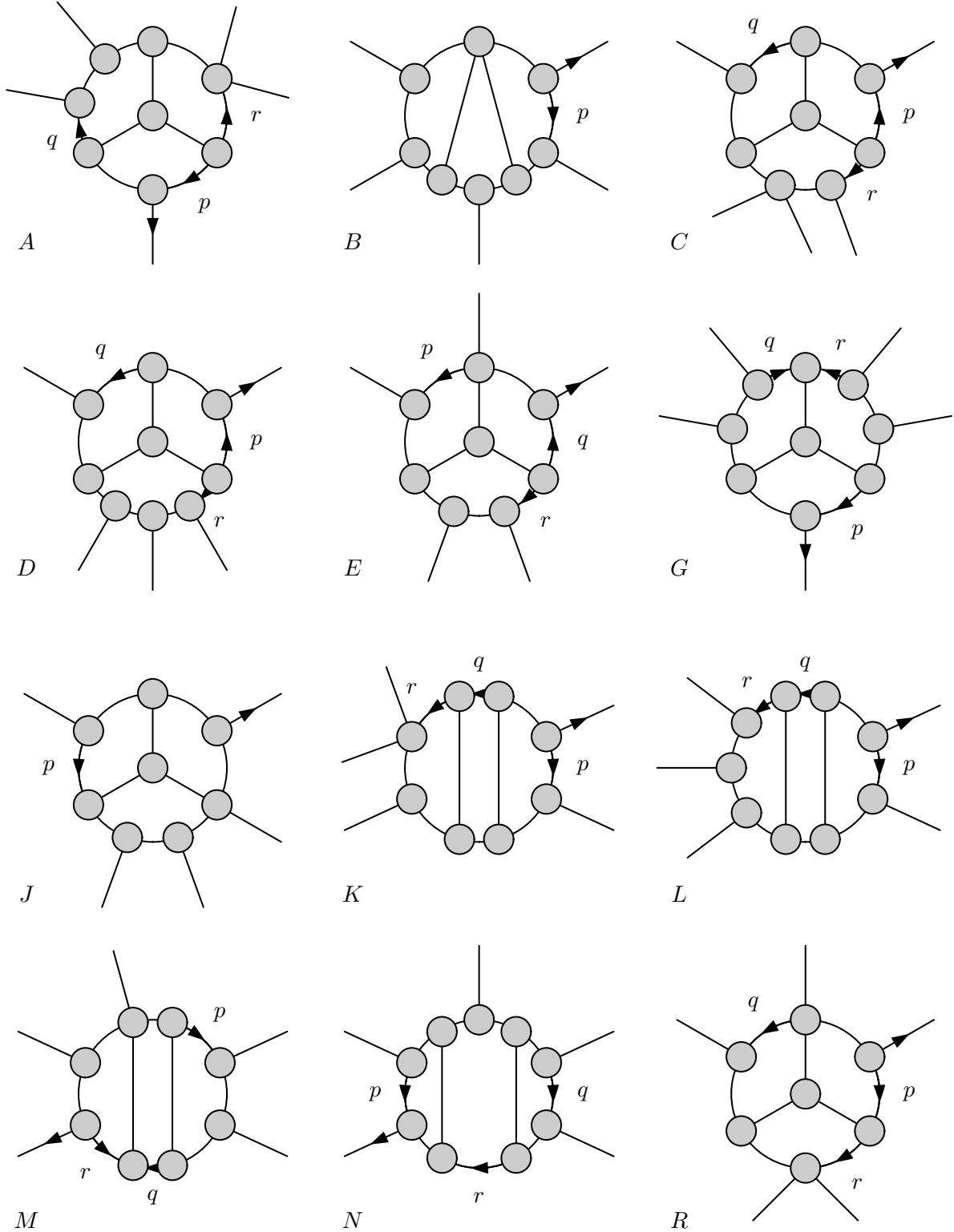}
\caption{The planar 3-loop 5-particle topologies
associated to leading singularities.  Each figure represents
a sum over that subset of Feynman diagrams in which all
of the indicated propagators are present.
We label the external momenta clockwise with
$k_1$ at the leg indicated with the arrow.}
\label{topologies}
\end{figure}

In order to proceed systematically we begin by enumerating all
planar 3-loop 5-particle topologies which are free of tadpoles, bubbles,
and triangles, since such diagrams  are unneccesary due to ${\mathcal{N}}=4$
supersymmetry (see~\cite{Bern:2006ew} for a thorough discussion).
This leaves 17 topologies, of which 5 do not have
any associated leading singularities and are therefore of no interest to us.
The remaining 12 topologies
are shown in~\fig{topologies}.

Each topology in~\fig{topologies} has several distinct
associated leading singularities, each of which gives rise to an equation
via~(\ref{eq:method}).  The information contained in this collection
of equations
is highly redundant---the equations obtained from only a small subset
of the leading singularities are sufficient to determine all coefficients,
while the remaining equations serve as consistency checks.
We now present a few details explaining how to extract a
set of equations sufficient for determining all coefficients.
We have verified a number of the additional equations to check consistency,
but have not performed an exhaustive search for all possible leading
singularities.

The topologies fall naturally into three different categories
according to how we choose to implement the collapse and expand rules.
Let us now address each category in turn, giving in each case the
details of the simplest topology as an example.

\subsubsection{Example 1: Topology $L$}

Topology $L$ has several leading singularities, but the simplest
ones can be isolated as indicated in the following cartoon:
\begin{eqnarray}
&&
{\hbox{\lower 26.pt\hbox{
\includegraphics{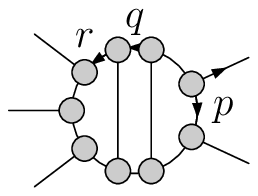}
}}}
{\int dp \atop \longrightarrow}
{\hbox{\lower 27.pt\hbox{
\includegraphics{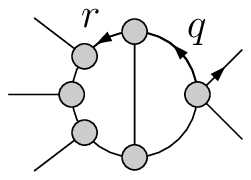}
}}}
\cr
&&
=
{\hbox{\lower 22.pt\hbox{
\includegraphics{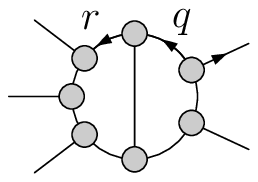}
}}}
 \!\!\!\!\!\!\! + {\cal O}([(q + k_1)^2]^0)
 ~~~
{\int dq \atop \longrightarrow}
{\hbox{\lower 26.pt\hbox{
\includegraphics{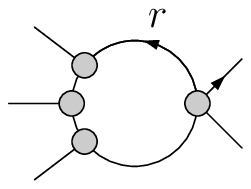}
\label{eq:exampleL}
}}}
\end{eqnarray}
In words: we first integrate the sum of
Feynman diagrams over a $p$ contour which collapses
the associated massless box,
then expand around $(q + k_1)^2 = 0$ keeping only the singular
terms indicated.  Integrating $q$ over an appropriate
contour isolates these singular terms while collapsing the massless box.
The final integral over $r$ is again accomplished using the collapse rule.

The leading singularities exposed by these steps are those
located at the locus
\begin{eqnarray}
S_L = \{ (p,q,r) \in {\mathbb{C}}^{12} &:&
p^2 = 0,~q^2 = 0,~r^2 = 0,~(p + k_1)^2 = 0,~
(r - k_5)^2 = 0,\cr
&&(r - k_{45})^2 = 0,~
(r + k_{12})^2 = 0,~
(q + k_{12})^2 = 0,~
(p - k_2)^2 = 0,\cr
&&(q - r)^2 = 0,~
(p + q + k_1)^2 = 0,~(q + k_1)^2 = 0\}.
\end{eqnarray}
For generic external momenta the set $S_L$ consists of 8 distinct points
in ${\mathbb{C}}^{12}$.  For each point in $S_L$ there is an associated
contour which computes the residue at the point and hence leads
to an equation via~(\ref{eq:method}).

It remains only to construct an appropriate ansatz for the left-hand
side of~(\ref{eq:method}).  We try a linear combination of the two
most natural integrals of topology $L$,
\begin{equation}
L
{\hbox{\lower 21.5pt\hbox{
\includegraphics{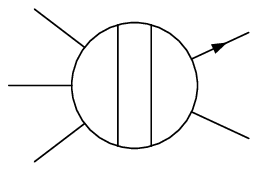}
}}}
+ L_1
{\hbox{\lower 21.5pt\hbox{
\includegraphics{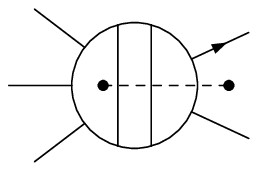}
}}}.
\label{eq:ansatz}
\end{equation}
Here and in what follows
we use pictures as shorthand for the corresponding scalar
integrands, so for example
the first term in~(\ref{eq:ansatz}) represents
\begin{equation}
L\,
\frac{1}{p^2 q^2 r^2 (p + k_1)^2
(r - k_5)^2 (r - k_{45})^2  (r + k_{12})^2 (q + k_{12})^2
(p - k_2)^2 (q - r)^2 (p + q + k_1)^2}
\end{equation}
while the dotted line in the second picture in~(\ref{eq:ansatz}) indicates
a factor of $(r + k_1)^2$ in the numerator of the integrand.

Integrating~(\ref{eq:ansatz}) over the contours detailed above
leads to the expression
\begin{equation}
{L + L_1 (r + k_1)^2 \over
s_{12}^2 s_{34} s_{45} (r + k_1)^2},
\end{equation}
where the denominator factors arise from the Jacobians in~(\ref{eq:jacobian}).
Equating this to the result of~(\ref{eq:exampleL}) and choosing
a particular helicity configuration leads to the equation
\begin{equation}
{L + L_1 (r + k_1)^2 \over  s_{12}^2 s_{34} s_{45} (r + k_1)^2}
=
A_{\rm tree}(1^-,2^-,3^+,4^+,5^+)\,\delta_{\langle r, 5 \rangle}.
\end{equation}
Of course this must be evaluated on the locus $S_L$, and it is easy
to check that $S_L$ contains only two different values of $r$:
\begin{equation}
r_1=\lambda_5\left(\widetilde{\lambda}_5+\frac{\langle 4,3\rangle}{\langle
5,3\rangle}\widetilde{\lambda}_4\right),
\qquad r_2=\left(\lambda_5+\frac{[4,3]}{[
5,3]}\lambda_4\right)\widetilde{\lambda}_5,
\end{equation}
giving us two distinct equations which are sufficient to determine
the coefficients $L$ and $L_1$ uniquely.

Topologies $D$, $G$, and $N$ proceed in exactly the same manner, except
that in these cases more than two integrals appear on the left-hand
side.
Topologies $A$, $C$, $E$ and $K$ are also very similar, except that since
these three topologies only have 10 exposed propagators (rather than 11)
it is necessary to isolate a second hidden
singularity by performing
a second expansion prior to integrating over $r$.

\subsubsection{Example 2: Topology $M$}

For topology $M$ it is sufficient to consider
even simpler contours.
We first collapse and expand the $p$ box as done above for topology $L$,
arriving at
\begin{equation}
{\hbox{\lower30.5pt\hbox{
\includegraphics{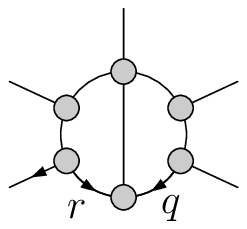}
}}}
\end{equation}
At this stage it is convenient to integrate over a symmetric contour
of the type considered in~\cite{Cachazo:2008vp}
where we require that $\langle q,r \rangle$ and $[q,r]$ separately
vanish instead of just $(q + r)^2 = 0$.
This leads us to consider the locus
\begin{eqnarray}
S_M = \{ (p, q, r) \in {\mathbb{C}}^{12} &:&
p^2 = 0,~q^2 = 0,~r^2 = 0,~(p-k_4)^2 = 0,~(r+k_1)^2 = 0,\cr
&& (q+k_{45})^2 = 0,~(r+k_{12})^2 = 0,~(p-q-k_{45})^2 = 0,\cr
&& (p-k_{45})^2 = 0,~(q+k_5)^2 = 0,~\langle q,r\rangle = 0,~
[q,r]= 0\}.
\end{eqnarray}
For generic external momenta $S_M$ consists of 4 isolated points, each
of which leads to one linear equation for the integral coefficients.
Note that the right-hand side of~(\ref{eq:method}) always vanishes
for such symmetric contours since the associated product of tree
amplitudes must vanish when $\langle q,r\rangle = 0 = [q,r]$.

For topologies $B$, $J$ and $G$
we proceed along exactly the same lines (we already treated $G$ in the
first example, but
additional equations
are needed to fix all of the coefficients which appear for this topology).
It turns out that for topology $J$ an interesting and very useful feature
emerges:
here we model the left-hand side as the linear combination
\begin{equation}
J
{\hbox{\lower 28.625pt\hbox{
\includegraphics{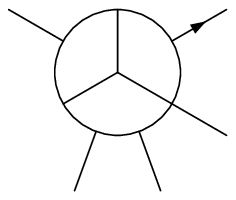}
}}}
+
J_1 {\hbox{\lower 28.625pt\hbox{
\includegraphics{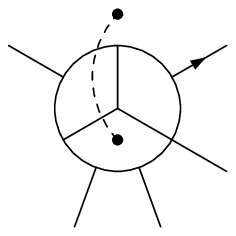}
}}} + {\rm several~other~integrals}.
\end{equation}
We will not display all of the relevant integrals explicitly,
but they all have the property that they either vanish on the locus
$S_J$ (so that they do not enter the associated equations), or they
contain the same numerator factor as $J_1$ shown here, which we denote
by $\ell^2$.  Now when we perform the $q$ integral one of the Jacobian
factors is $1/\ell^2$, so we obtain the equation
\begin{equation}
\label{eq:more}
{J \over \ell^2} + J_1 + {\rm several~other~coefficients} = 0.
\end{equation}
Since $\ell^2 = 0$ on the locus $S_J$, we immediately see that the
coefficient $J$ must vanish in order to avoid a contradiction.
Perhaps a safer way to express this is to say that we can consider
an equation obtained by multiplying both sides of~(\ref{eq:method})
by $\ell^2$ before performing the contour integrals.
Having determined that $J = 0$, we then see that~(\ref{eq:more})
gives an equation relating $J_1$ to the other coefficients.
This trick is also useful for other topologies, in particular
for $C$ and $E$.

\subsubsection{Example 3: Topology $R$}

There are five different triple-box 9-propagator
topologies, of which topology $R$ is the only one with
associated leading singularities.  These are situated on the locus
\begin{eqnarray}
S_R = \{ (p,q,r) \in {\mathbb{C}}^{12} &:&
p^2 = 0,~(p + q + k_{15})^2 = 0,~ q^2 = 0,~(q-k_4)^2 = 0,~(p-r)^2 = 0,
\cr
&&\qquad
(r - k_{23})^2 = 0,~r^2 = 0,~(r + q + k_{15})^2 = 0,~(p+k_1)^2 = 0,
\cr
&&\qquad
(q + k_{15})^2 = 0,~(r + k_1)^2 = 0,~(r + k_{15})^2 = 0 \}.
\label{eq:Rleading}
\end{eqnarray}
Here the first nine conditions are the visible propagators,
while the last three are hidden singularities.
In order to see what the right-hand side of~(\ref{eq:method})
should be let us begin by integrating out $q$ to collapse the first
box.  This leads to
\begin{equation}
{\hbox{\lower 31.25pt\hbox{
\includegraphics{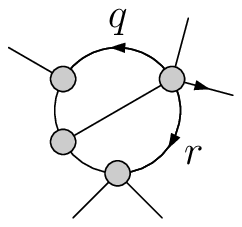}
}}}
\label{eq:toexpose}
\end{equation}
For the first time we find a triangle-triangle diagram rather
than a triangle-box or box-box.
The Jacobian factor from integrating the corresponding scalar
integral is $1/(q+ k_{15})^2 (r + k_1)^2$, suggesting that
we expand~(\ref{eq:toexpose}) to expose either
$1/(q + k_{15})^2$ or the $1/(r + k_1)^2$
propagator, but it is clearly impossible
to expand both simultaneously.  Either choice leaves us with a
sum over triangle-box Feynman diagrams, which vanishes
due to ${\mathcal{N}} = 4$
supersymmetry.
The right-hand side of~(\ref{eq:method}) is therefore zero for
the $R$ topology leading singularities in eq.~(\ref{eq:Rleading}).

\section{The 3-loop 5-particle Amplitude}

\begin{figure}
\includegraphics{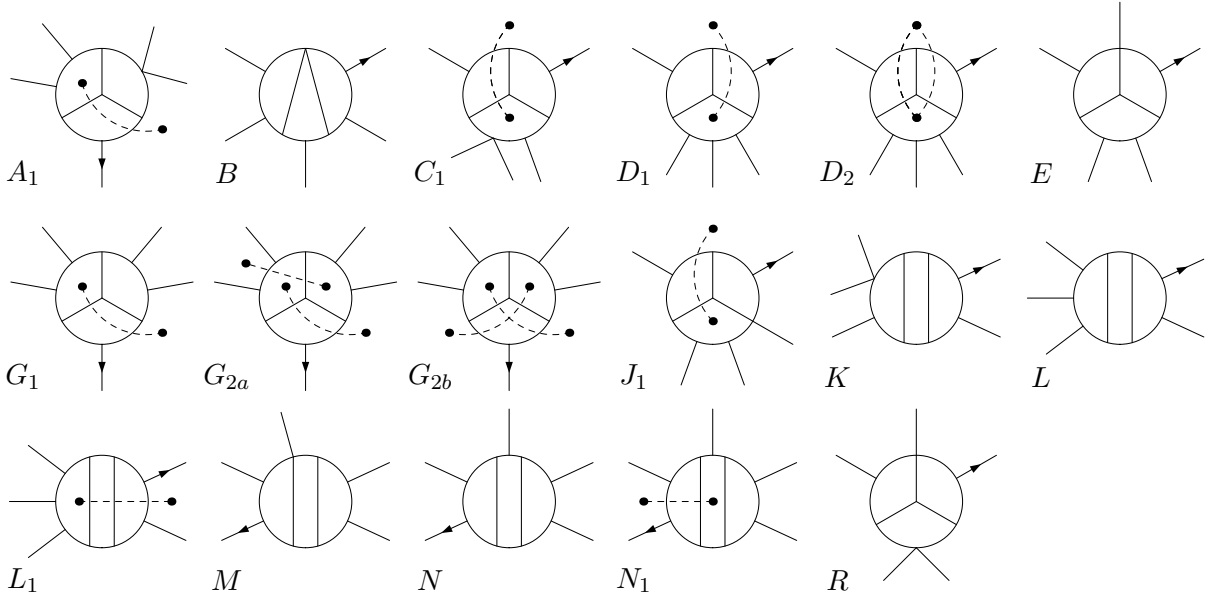}
\caption{The 17 independent integrals appearing in the ansatz.  Other
integrals can be obtained by rotations or reflections. As
in~\fig{topologies} we label the external momenta clockwise
with $k_1$ at the position indicated by the arrow.}
\label{basis}
\end{figure}

A basis of integrals which is sufficient for representing
all of the leading singularities of the amplitude is
shown in~\fig{basis}.
By solving the collection of linear equations
as explained in the previous section we find their coefficients
\begin{eqnarray}
A_1 &=& - s_{12}
s_{23}^2 {1- \gamma_3 \over \gamma_3 - \widetilde{\gamma}_3}, \cr B
&=& - s_{12} s_{23} s_{34} s_{45} {\gamma_5 \over \gamma_5 -
\widetilde{\gamma}_5},\cr C_1 &=& -s_{12} s_{51}^2
{\widetilde{\gamma_3} \over \gamma_3 -\widetilde{\gamma}_3}, \cr D_1
&=& s_{12} s_{23} s_{51}^2 {1 \over \gamma_3 -\widetilde{\gamma}_3},
\cr D_2 &=& - s_{23} s_{34} s_{15} {1- \gamma_4 \over \gamma_4 -
\widetilde{\gamma}_4},\cr E &=& - s_{12} s_{23}^2 s_{51} {1 \over
\gamma_3 -\widetilde{\gamma}_3}, \cr G_{1} &=& s_{12} s_{23}^2 s_{51}
{1 \over \gamma_3 -\widetilde{\gamma}_3}, \cr G_{2a} &=&
- s_{23} s_{45} s_{51} {\widetilde{\gamma}_2 \over \gamma_2 -
\widetilde{\gamma}_2}, \cr G_{2b} &=& -s_{23} s_{34} s_{45} {1 \over
\gamma_5 -\widetilde{\gamma}_5}, \cr J_1 &=&  - s_{34} s_{45} s_{51}
{1 \over \gamma_1 - \widetilde{\gamma}_1},\cr K &=& - s_{12}^3
s_{23} {1- \gamma_3 \over \gamma_3 - \widetilde{\gamma}_3}, \cr L
&=& s_{12}^3 s_{23} s_{51} {1 \over \gamma_3 -
\widetilde{\gamma}_3},\cr L_1 &=& - s_{12}^2 s_{34} s_{45}
{\widetilde{\gamma}_1 \over \gamma_1 - \widetilde{\gamma}_1},\cr M
&=& - s_{12} s_{45}^2 s_{51} {1 \over \gamma_2 -
\widetilde{\gamma}_2},\cr N &=& s_{51} s_{12} s_{34} s_{45}^2 {1
\over \gamma_1 - \widetilde{\gamma}_1},\cr N_1 &=& - s_{12} s_{34}
s_{45}^2 {\widetilde{\gamma}_1 \over \gamma_1 -
\widetilde{\gamma}_1},\cr R &=& s_{23} s_{45} s_{51}
{1-\gamma_1 \over \gamma_1 -\widetilde{\gamma}_1}
\label{eq:coefs}
\end{eqnarray}
where we have introduced the quantity
\begin{equation}
\gamma_i=\left(1+\frac{\langle i+2,i+3\rangle[i+3,i]}{\langle
i+2,i+4\rangle[i+4,i]}\right)^{-1},
\end{equation}
and $\widetilde{\gamma}$ is given by the parity conjugate
of this expression (i.e., $\langle a,b\rangle \leftrightarrow [a,b]$).
In each case we have suppressed an overall factor of $A_5^{\rm tree}$.
In order to connect to the standard normalization conventions used
in the study of the BDS ansatz it is necessary to multiply by an overall
factor of $(-1/2)^L$.
The complete amplitude is therefore assembled via the formula
\begin{equation}
\label{eq:assembly}
M^{(3)}_5 = A^{(3)}_5/A^{\rm tree}_{5}
= - \frac{1}{8} \sum_{{\rm permutations}}
\sum_{{\rm integrals}}
\frac{1}{S_i} {\rm coefficient}_i \times {\rm integral}_i.
\end{equation}
The first sum runs over the 10 cyclic and anti-cyclic orderings
of the labels $1,2,3,4,5$ of the external particles
and the second sum runs over
the 17 integrals in~\fig{basis}.  $S_i$ is a symmetry factor
to compensate for possible overcounting: $S = 2$ for
integrals $B$, $D_1$, $D_2$, $G_{2b}$, $L$, $L_1$, $N$, and $R$,
and $S=1$ for the others.

The presentation~(\ref{eq:coefs}) makes it simple to read off
the parity-even parts of the coefficients, which will be useful in the
following section.  We find the parity-even parts
\begin{eqnarray}
\label{eq:evenpart} A_1 = \frac{1}{2} s_{12} s_{23}^2, \qquad &&
\qquad B = - \frac{1}{2} s_{12} s_{23} s_{34} s_{45}, \cr C_1 =
\frac{1}{2} s_{12} s_{51}^2, \qquad && \qquad D_2 = \frac{1}{2}
s_{23} s_{34} s_{51}, \cr G_{2a} = \frac{1}{2} s_{23} s_{45}
s_{51},\qquad && \qquad K = \frac{1}{2} s_{12}^3 s_{23},\cr L_1 =
\frac{1}{2} s_{12}^2 s_{34} s_{45},\qquad && \qquad N_1 =
\frac{1}{2} s_{12} s_{34} s_{45}^2, \cr R &=& - \frac{1}{2} s_{23}
s_{45} s_{51}.
\end{eqnarray}
with all others vanishing.
We note that only the coefficients associated
to dual conformal integrals have non-vanishing parity-even parts, as expected
based on the pattern of previously studied amplitudes~\cite{BDS,TwoLoopFiveA,TwoLoopFiveB,Bern:2006ew,Bern:2008ap}.

\section{The Three-Loop BDS Ansatz}

The infrared divergences of higher loop scattering amplitudes in gauge
theory are very simply related to those of lower loop
amplitudes~\cite{KnownIR}.
In~\cite{ABDK,BDS}, it was conjectured
that in ${\mathcal{N}} = 4$ SYM
this simplicity persists, at least for MHV amplitudes, to the
finite terms as well.
Although the explicit $n=6$ calculation of~\cite{Bern:2008ap}
has now demonstrated, following earlier doubts raised
in~\cite{AMTrouble,BNST,Lipatov} (see also~\cite{Bartels:2008sc}), that
these relations are not true for all $n$, it
is believed that they should hold for four
and five particles at any number of loops since for these cases the
amplitudes are determined up to a few constants
by dual conformal
invariance~\cite{AMTrouble,ConformalWard}.

The precise form of the BDS ansatz
at three loops, in dimensional regularization to $D = 4 - 2
\epsilon$, is
\begin{equation}
\label{eq:bds}
M_n^{(3)}(\epsilon) = - \frac{1}{3} (M_n^{(1)}(\epsilon))^3 +
M_n^{(1)}(\epsilon) M_n^{(2)}(\epsilon) + f^{(3)}(\epsilon)
M_n^{(1)}(3 \epsilon) +
C^{(3)} + {\cal O}(\epsilon)
\end{equation}
where
\begin{equation}
f^{(3)}(\epsilon) = \frac{11 \pi^4}{180} + \left(5 \zeta(2)\zeta(3)
+ 6 \zeta(5) \right) \epsilon +  a \epsilon^2, \qquad
C^{(3)} = b
\end{equation}
in terms of two previously
undetermined numerical constants $a$ and $b$.
BDS verified by explicit calculation that the 3-loop 4-particle
amplitude satisfies~(\ref{eq:bds}), but the structure
of the equation for $n=4$ is insensitive to the values of $a$ and $b$
as long as they obey the linear relation
\begin{equation}
\label{eq:linearone}
2 a - 9 b = - \frac{341}{24} \zeta(6) + 17 \zeta(3)^2.
\end{equation}
Here we will use our 3-loop 5-particle amplitude to extract a second
linear equation from~(\ref{eq:bds}) which will finally fix the
constants $a$ and $b$.

The calculation of $a$ and $b$ benefits from two simplifcations.
The first is that we may restrict our attention to
the parity-even part of~(\ref{eq:bds}).
If we write each amplitude as a sum of its parity-even
and parity-odd parts, $M_5^{(L)} = M_{5+}^{(L)} + M_{5-}^{(L)}$, then
taking the parity-even part of~(\ref{eq:bds}) for $n=5$ gives
\begin{eqnarray}
\label{eq:simpbds}
M_{5+}^{(3)}(\epsilon) &=& - \frac{1}{3} (M_{5+}^{(1)}(\epsilon))^3 +
M_{5+}^{(1)}(\epsilon) M_{5+}^{(2)}(\epsilon) + f^{(3)}(\epsilon)
M_{5+}^{(1)}(3 \epsilon) +
C^{(3)}\cr
&& + M_{5-}^{(1)}(\epsilon) \left(
M_{5-}^{(2)}(\epsilon)  - M_{5+}^{(1)}(\epsilon)  M_{5-}^{(1)}(\epsilon)
\right) + {\cal O}(\epsilon).
\end{eqnarray}
In~\cite{TwoLoopFiveB} it was shown that
$M_{5-}^{(2)}(\epsilon)  - M_{5+}^{(1)}(\epsilon)  M_{5-}^{(1)}(\epsilon)
= {\cal O}(\epsilon)$.
Since $M_{5-}^{(1)}(\epsilon)$
itself is also ${\cal O}(\epsilon)$, we see that
the entire last line of~(\ref{eq:simpbds}) can be replaced simply by
$+{\cal O}(\epsilon)$.  A consequence of the result of~\cite{TwoLoopFiveB}
is therefore that the parity-even part of the three-loop BDS ansatz
can be obtained
by making the naive replacement $M_5^{(3)} \to M_{5+}^{(3)}$
in~(\ref{eq:bds}).

The second simplication is to make use of the notion of
obstructions introduced in~\cite{Cachazo:2006mq} and exploited in the
four-loop calculations~\cite{Cachazo:2006az,Cachazo:2007ad}.
We refer the reader to~\cite{Cachazo:2006az} for all of the necessary details,
including a detailed algorithm for calculating obstructions.  Here
we simply remind the reader that for an amplitude $A(x,\epsilon)$
depending on a single
kinematic variable $x$, the obstruction $P(\epsilon)$
is defined
to be the coefficient of the simple pole at $y=0$ in the inverse
Mellin transform
transform, so that
\begin{equation}
\label{eq:obstructions}
A(x,\epsilon) = \int_{-i \infty}^{+i \infty} dy\
x^y \left[ ({\rm higher~order~singularities}) +  \frac{P(\epsilon)}{y}
+ ({\rm regular~at~} y=0)
\right].
\end{equation}
As explained in~\cite{Cachazo:2006az} it is important to understand
this relation as holding order by order in $\epsilon$, rather than
at finite $\epsilon$.
The prime advantage of dealing with $P(\epsilon)$
rather than the full $A(x,\epsilon)$ is that it is much simpler to
compute.  Furthermore it is important that obstructions satisfy
a product algebra---the obstruction in any product of amplitudes
is equal to the product of the individual obstructions.

This concept generalizes straightforwardly to integrals depending
on more than one kinematic variable.  In the case at hand we
have 5-particle integrals depending on five independent variables
$s_{i,i+1}$, and we can extract the obstruction $P(\epsilon)$ by
applying the above procedure five times in succession.
Equivalently, we define $P(\epsilon)$ to be the coefficient of the
$1/(y_1 y_2 y_3 y_4 y_5)$ pole in the 5-fold inverse Mellin transform
of $A(s_{12},s_{23},s_{34},s_{45},s_{51})$.
By applying the algorithm outlined in~\cite{Cachazo:2006az} it
is straightforward to find that
the obstructions in the one- and two-loop five particle amplitudes are
given by
\begin{eqnarray}
\label{eq:obsonetwo}
P_5^{(1)}(\epsilon) &=& - \frac{5}{2} \frac{1}{\epsilon^2} + \frac{5 \pi^2}{8} +
\frac{179 \zeta(3)}{24} \epsilon + \frac{97 \pi^4}{1440} \epsilon^2
- \left( \frac{51 \pi^2 \zeta(3)}{32} - \frac{137 \zeta(5)}{8}
\right) \epsilon^3\cr
&&\qquad- \left( \frac{763 \zeta(3)^2}{72} - \frac{23
\pi^6}{3780} \right) \epsilon^4 + {\cal O}(\epsilon^5), \cr
P_5^{(2)}(\epsilon) &=& \frac{25}{8}
\frac{1}{\epsilon^4} - \frac{35 \pi^2}{24} \frac{1}{\epsilon^2} -
\frac{865 \zeta(3)}{48} \frac{1}{\epsilon} - \frac{97 \pi^4}{1152} +
21.494969  \epsilon
\cr
&&\qquad
- 64.357473 \epsilon^2
+ {\cal O}(\epsilon^3).
\end{eqnarray}
For simplicity we have restricted $P_5^{(L)}$
here to the parity-even parts of the
amplitudes.
Note that these expressions satisfy the two-loop ABDK relation
\begin{equation}
P_5^{(2)}(\epsilon) = \frac{1}{2} (P_5^{(1)}(\epsilon))^2
+ ( - \zeta(2) - \zeta(3) \epsilon - \zeta(4) \epsilon^2)
P_5^{(1)}(2 \epsilon) - \frac{\pi^4}{72} + {\cal O}(\epsilon)
\end{equation}
as expected.

At three loops, we have found that there are nine independent integrals
which contribute to the parity-even part of the 5-particle amplitude.
The obstructions for each of these types of integrals,
through ${\cal O}(\epsilon^0)$, are
\begin{eqnarray}
P_{A_1} &=& \frac{20}{9} \frac{1}{\epsilon^6} + \frac{20\pi^2}{27}
\frac{1}{\epsilon^4} - \frac{43 \zeta(3)}{2} \frac{1}{\epsilon^3} +
\frac{73 \pi^4}{432} \frac{1}{\epsilon^2} -
850.242028
 \frac{1}{\epsilon}
 + 34.239832,
\cr P_B &=& \frac{70}{3} \frac{1}{\epsilon^6} - \frac{45 \pi^2}{2}
\frac{1}{\epsilon^4} - \frac{1495 \zeta(3)}{6} \frac{1}{\epsilon^3}
- \frac{76 \pi^4}{135} \frac{1}{\epsilon^2} +
1589.962798
\frac{1}{\epsilon} + 2824.770745,
\cr P_{C_1} &=& \frac{20}{9} \frac{1}{\epsilon^6} -
\frac{25 \pi^2}{54} \frac{1}{\epsilon^4} - \frac{557 \zeta(3)}{36}
\frac{1}{\epsilon^3} + \frac{17137 \pi^4}{12960}
\frac{1}{\epsilon^2} +
221.894995 \frac{1}{\epsilon} + 1030.164974, \cr P_{D_2} &=&
\frac{35}{3} \frac{1}{\epsilon^6} - \frac{355 \pi^2}{36}
\frac{1}{\epsilon^4} - \frac{645 \zeta(3)}{4} \frac{1}{\epsilon^3} +
\frac{767 \pi^4}{2160} \frac{1}{\epsilon^2} +
231.123687
\frac{1}{\epsilon} -4141.657880,
\cr P_{G2a} &=& 20 \frac{1}{\epsilon^6} - \frac{155
\pi^2}{9} \frac{1}{\epsilon^4} - \frac{563 \zeta(3)}{4}
\frac{1}{\epsilon^3} - \frac{487 \pi^4}{288} \frac{1}{\epsilon^2} +
1294.520402 \frac{1}{\epsilon} + 2938.6610 \pm 0.0036,
\cr P_{K} &=& \frac{20}{9}
\frac{1}{\epsilon^6} - \frac{5 \pi^2}{18} \frac{1}{\epsilon^4} -
\frac{1177 \zeta(3)}{36} \frac{1}{\epsilon^3} + \frac{719
\pi^4}{4320} \frac{1}{\epsilon^2} +
178.487460
 \frac{1}{\epsilon}
 -2387.290195,
\cr P_{L_1} &=& \frac{35}{3} \frac{1}{\epsilon^6} - \frac{85 \pi^2}{12}
\frac{1}{\epsilon^4} - \frac{1411 \zeta(3)}{12} \frac{1}{\epsilon^3}
- \frac{1195 \pi^4}{432} \frac{1}{\epsilon^2}
-673.319831 \frac{1}{\epsilon}
- 2845.889639,
\cr P_{N_1} &=& 15 \frac{1}{\epsilon^6} - \frac{455
\pi^2}{36} \frac{1}{\epsilon^4} - 136 \zeta(3) \frac{1}{\epsilon^3}
+ \frac{983 \pi^4}{1440} \frac{1}{\epsilon^2} +
625.875398
\frac{1}{\epsilon} + 437.509754 \cr P_R &=& - \frac{80 \zeta(3)}{3}
\frac{1}{\epsilon^3} + \frac{107 \pi^4}{108} \frac{1}{\epsilon^2} -
395.562804
\frac{1}{\epsilon}
+ 923.415196.
\end{eqnarray}
Each expression displays the result obtained after summing over all
10 permutations of the corresponding integral (including in each
case the appropriate
dual conformal numerator).
The estimated error in the numerical results is much smaller
than the precision indicated in all cases except for the last
term in $P_{G2a}$, which is the overwhelmingly dominant source of
numerical error.

Using the parity-even parts of the coefficients obtained in the previous
section, and including the necessary factors of $1/2$ to avoid overcounting
those integrals with flip symmetries, we find the total three-loop
obstruction
\begin{eqnarray}
\label{eq:threeloopobs}
P_5^{(3)} &=& - \frac{1}{16} \left( P_{A_1} - \frac{1}{2} P_B + P_{C_1} +
\frac{1}{2} P_{D_2} +
P_{G2a} + P_K + \frac{1}{2} P_{L_1} + P_{N_1} - \frac{1}{2} P_R \right)
\cr &=&-\frac{125}{48} \frac{1}{\epsilon^6} + \frac{325 \pi^2}{192}
\frac{1}{\epsilon^4} + \frac{4175 \zeta(3)}{192}
\frac{1}{\epsilon^3} + \frac{499 \pi^4}{10368} \frac{1}{\epsilon^2}
\cr
&&
- 40.764885
 \frac{1}{\epsilon}
+ 207.1613 \pm 0.0002
+ {\cal O}(\epsilon)
\end{eqnarray}
Using the results~(\ref{eq:threeloopobs}) and~(\ref{eq:obsonetwo}),
we find that the BDS relation~(\ref{eq:bds}) is satisfied provided that
$a$ and $b$ satisfy the linear relation
\begin{equation}
5 a - 18 b = 105.482 \pm 0.004.
\end{equation}
Together with~(\ref{eq:linearone}) this implies the solution
\begin{equation}
a = 85.263 \pm 0.004, \qquad
b = 17.8241 \pm 0.0009.
\end{equation}
Based on the transcendentality hypothesis, it is expected
that each of these numbers should be a linear combination of
$\zeta(6)$ and $\zeta(3)^2$ with rational
coefficients.  However given the
limited numerical accuracy of our calculation it seems
prudent to avoid speculating on possible exact values for $a$ and
$b$ at this time.

\section{Summary}

In this paper we have used the leading singularity method
to obtain an ansatz for the four-dimensional
cut-constructible part of the 3-loop 5-particle amplitude
in ${\mathcal{N}} = 4$ SYM theory.
This means that we have determined the coefficients of the
integrals shown in~\fig{basis} by comparing residues of the ansatz
to those of the amplitude on various leading singularities.
Although
it has not yet been proven that determination of only leading
singularities completely determines an amplitude (in principle
one might have to add additional integrals that vanish on all leading
singularities but that have subleading singularities), the method
so far has been found to give the complete answer in all cases
where comparison with alternate methods was possible.

Dimensionally regulated amplitudes occasionally contain
so-called `$\mu$-terms' which are defined as terms in the integrand
which vanish in $D=4$ but not in $D = 4 - 2 \epsilon$ (note that
this statement is, in general, completely unrelated to whether or
not these terms vanish in $D = 4 - 2 \epsilon$ after integration;
indeed $\mu$-terms can easily be IR divergent).
Since the leading singularity method itself works
with strictly four-dimensional loop momenta, it is insensitive
to possible $\mu$ terms, although it seems that in principle
they could be determined by considering leading singularities
in integer dimensions other than 4.  However,
in all cases that have been studied so far it has been observed
that $\mu$-terms separately cancel out of the BDS relation, leaving
$C^{(3)}$ unaffected.
We can therefore hope that even if the 3-loop 5-particle amplitude
contains such terms which we have missed, they would not contribute
to the constants $a$ and $b$ computed
in section~IV.

Finally we emphasize that since our goal in section~IV was to streamline
the calculation of $a$ and $b$ as much as possible, we have only
evaluated the obstructions, not the full amplitude.
Consequently we have not checked (even numerically) that the
quantity $+ C^{(3)}$ appearing in~(\ref{eq:method}) is a numerical
constant; in principle it could depend on the kinematic
variables $s_{i,i+1}$.
The method of obstructions is efficient
for quickly extracting the `constant part' of $C^{(3)}$
(defined as the coefficient of $1/y$ in the inverse Mellin transform)
but is insensitive to any other potential terms in $C^{(3)}$ that
depend on the $s_{i,i+1}$.
It remains an interesting open problem to
verify that there are no such terms.

\section*{Acknowledgments}

We have benefited from from useful discussions
with Z.~Bern, F.~Cachazo, L.~Dixon, A.~Jevicki,
D.~Kosower, R.~Roiban and C.-I. Tan.
M. S. is grateful to Banff International Research Station
and to the Tata Institute for Fundamental Research, and
A. V. to the ETH Z\"urich,
for kind hospitality and support during the course of this work.
This work was supported in part by the
US Department of Energy under contract
DE-FG02-91ER40688 (M. S. (OJI) and A. V.), and the
US National Science Foundation under grants PHY-0638520 (M. S.) and
PHY-0643150
CAREER (A. V.).

\end{document}